\documentclass[preprint,showpacs,preprintnumbers,amsmath,amssymb]{revtex4}%
\usepackage{graphicx}
\usepackage{dcolumn}
\usepackage{bm}
\usepackage{amsmath}
\usepackage{amsfonts}
\usepackage{amssymb}%

\begin{document}

\title{Classically and Quantum stable  Emergent Universe from Conservation Laws}

\author{Sergio del Campo\footnote{Deceased}}
\affiliation{ Instituto de F\'{\i}sica, Pontificia
Universidad Cat\'{o}lica de Valpara\'{\i}so, Avenida Brasil 2950,
Casilla 4059, Valpara\'{\i}so, Chile.}
\author{Eduardo I. Guendelman}
\email{guendel@bgu.ac.il} \affiliation{ Physics Department, Ben
Gurion University of the Negev, Beer Sheva 84105, Israel}
\author{Ram\'on Herrera}
\email{ramon.herrera@ucv.cl} \affiliation{ Instituto de
F\'{\i}sica, Pontificia Universidad Cat\'{o}lica de
Valpara\'{\i}so ,  Avenida Brasil 2950, Casilla 4059,
Valpara\'{\i}so, Chile.}
\author{Pedro Labra$\tilde{n}$a}
\email{plabrana@ubiobio.cl} \affiliation{ Departamento de
F\'{\i}sica, Universidad  del B\'{\i}o B\'{\i}o and Grupo de
Cosmolog\'{\i}a y Gravitaci\'on-UBB, Avenida Collao 1202, Casilla
5-C, Concepci\'on, Chile.}


\date{\today}

\begin{abstract}

It has been recently pointed out by Mithani-Vilenkin
\cite{Mithani:2014jva, Mithani:2011en, Mithani:2012ii,
Mithani:2014toa} that certain emergent universe scenarios which are
classically stable are nevertheless unstable semiclassically to
collapse. Here, we show that there is a class of emergent universes
derived from scale invariant two measures theories with spontaneous
symmetry breaking (s.s.b) of the scale invariance, which can have
both classical stability and do not suffer the instability pointed
out by Mithani-Vilenkin towards collapse.

We find that this stability is due to the presence of a symmetry in
the "emergent phase", which together with the non linearities of the
theory, does not allow that the FLRW scale factor to be smaller that
a certain minimum value $a_0$ in a certain protected region.

\end{abstract}

\maketitle


The standard hot big-bang model provides us with the description of
how the universe evolves, explaining  the observational facts, such
as the Hubble expansion, the $3 K$ microwave background radiation
and the abundance of light elements. However, this model presents
some problems in its evolution. We will reference some of then; the
smoothness or horizon problem, the flatness, the structure or
primordial density problem, etc.. These problems can be solved in
the context of the inflationary universe \cite{Guth}, where the
essential feature of any inflationary model is the rapid but finite
period of expansion that the universe underwent at very early times
in its evolution.  Perhaps the most import feature of the
inflationary universe model is that it provides a causal explication
for the origin of the observed anisotropy in the cosmic microwave
background radiation (CMB), and also to the distribution of
large-scale structures, which are consistent with the observations
\cite{Larson, Planck}.


However, one should point out that even in the context of the
inflationary scenario one still encounters the initial singularity
problem \cite{singular, HE-book} showing that the universe
necessarily had a singular beginning for generic inflationary
cosmologies \cite{Borde-Vilenkin-PRL}.

One interesting way to avoid the initial singularity problem is to
consider the emergent universe (EU) scenario \cite{orgemm}.
The emergent universe refers to models in which the universe emerges
from a past eternal Einstein static state (ES), inflates, and then
evolves into a hot big bang era.
The EU is an attractive scenario since it avoids the initial
singularity and provides a smooth transition towards an inflationary
period.

The original proposal for the emergent universe \cite{orgemm}
supported an instability at the classical level of the ES state, and
various models intended to formulate a stable model have been given
\cite{emm2}, in particular the Jordan Brans Dicke models
\cite{emm3}.
In this context, Mithani-Vilenkin in
Refs.~\cite{Mithani:2011en}-\cite{Mithani:2014toa} have shown  that
certain classically stable static universes could be unstable
semiclassically towards collapse.
In this work, we show that there is a class of emergent universes
derived from scale invariant two measures theories with spontaneous
symmetry breaking of the scale invariance, which can have both
classical stability and do not suffer the instability pointed out by
Mithani-Vilenkin towards collapse of the ES state.

In a series of papers \cite{prev}-\cite{Guendelman:2014bva} we have
studied a class of EU scenarios which are based on a spontaneously
broken scale symmetry induced by the dynamics of a Two Measures
Field Theory (TMT)\cite{GK1}-\cite{GK8}, (see also Ref.\cite{prev}).
In such model there is a dilaton field $\phi$ and the EU as the
$t\longrightarrow -\infty$ is well described by an Einstein static
universe, where $t$ is the cosmological time.

In the Appendix A, we have included a summary of the principal
characteristic of the TMT theories. Here we want to consider the
detailed analysis of the EU solutions of the model developed in
Ref.~\cite{yo}. The results obtained in this case can also be
applied to models studied in
Refs.~\cite{prev}-\cite{Guendelman:2014bva}, which present similar
symmetries as the model in Ref.~\cite{yo}.

We start by considering the Friedmann-Robertson-Walker  closed
cosmological solutions of the form
\begin{equation}
ds^2 =dt^2 - a(t)^2 \left(\frac{dr^2}{1 -r^2}+ r^2(d\theta^2
+sin^2\theta d\phi^2)\right),   \phi = \phi(t),
\end{equation}
where $a(t)$ is the scale factor, and the scalar field $\phi$ is a
function of the cosmic time $t$ only, due to homogeneously and
isotropy.
We will consider a scenario where the scalar field $\phi$ is moving
in the extreme left region $\phi \rightarrow -\infty  $. In this
case, the expressions for the energy density $\rho$ and pressure $p$
are given by
\begin{equation}\label{eq.density}
\rho = \frac{A}{2} \dot{\phi}^2 + 3B\dot{\phi}^4 + C,
\end{equation}
and
\begin{equation}
p = \frac{A}{2} \dot{\phi}^2 +B\dot{\phi}^4 - C,\label{presion}
\end{equation}
see Appendix A. Where the constants $A,B$ and $C$ are given by,
\begin{eqnarray}
A = 1- \frac{2\delta b_g V_1}{4(b_g V_1 - V_2)}\,,\,\,\,\, B =
-\frac{\delta^2 b^2_g }{4(b_g V_1 - V_2)}\,,\,\,\,\,
\mbox{and}\,\,\,\, C = \frac{ V^2_1}{4(b_g V_1 - V_2)}\,\label{C}.
\end{eqnarray}

As was discussed in Ref.\cite{yo}, the emergent universe can turn
into inflation only if $C>0$.
On the other hand, in order to have a scenario in which the emergent
universe evolves from an static and classically stable universe at
$a = a^*$ with
\begin{equation}\label{consta}
a^* = \sqrt{\left(\frac{3}{8\pi G}\right)\frac{12B}{A^2 + 24B\,C -
A\sqrt{A^2 + 12B\,C}}}\,\,,
\end{equation}
and then passed to an inflationary phase, the following conditions
must to be met, see Ref.~\cite{yo}:
\begin{eqnarray}
0.5<y<0.54\,, \label{ES1}\\
B < 0\,, \label{ES2}\\
-\frac{1}{64B} < C < -\sqrt{\frac{3}{B^2}} -
\frac{7}{4B}\,.\label{ES3}
\end{eqnarray}
Where $A=1-y$ and we have defined $y=\frac{2\delta b_g C}{V_1}$.

Now will turn our attention to possible quantum tunneling from the
solution $a = a^*$ to $a=0$, during the static regimen of the EU
scenario. Let us first note that there is a conserved quantity
$\Pi_\phi$, due to the fact that from $\phi\rightarrow -\infty$,
there is other symmetry $\phi\rightarrow \phi+ c$. Given that in the
Einstein frame we can use the action

\begin{equation}
S=\frac{1}{\kappa}\left[\int\,
R\,\sqrt{-g}\,d^4x\,+\,\int\,p\,\sqrt{-g}\,d^4x\right],
\end{equation}
and the symmetry $\phi\rightarrow \phi+ c$, in which c is a
constant, leads to the conservation law

\begin{equation}\label{Cons1}
a^3(t)\,[A\dot{\phi}+4\,B\dot{\phi}^3]=\Pi_\phi=const.\,.
\end{equation}


Without loss of generality, let us consider $\Pi_\phi
> 0$, see Appendix B for the case $\Pi_\phi
< 0$.
From conservation equation (\ref{Cons1}), we can write $a$ as a
function of $\dot{\phi}$

\begin{equation}\label{Cons1a}
a(\dot{\phi}) =
\left(\frac{\Pi_\phi}{A\dot{\phi}+4\,B\dot{\phi}^3}\right)^{1/3} .
\end{equation}

We can note that in this case $-\infty < \dot{\phi}<
-\sqrt{\frac{A}{4|B|}}$  or $0 <\dot{\phi}< \sqrt{\frac{A}{4|B|}}$
in order to satisfied $\Pi_\phi
> 0$.
When $\dot{\phi}$ is in the first region $a(\dot{\phi})$ is a
function which approach to zero when $\dot{\phi} \rightarrow -
\infty$ and diverges when $\dot{\phi} \rightarrow
-\sqrt{\frac{A}{4|B|}}$. But in this region $\rho$ becomes negative
see Eq.~(\ref{eq.density}), then we are not interested in this case.

On the other hand, when $\dot{\phi}$ is in the second region,
$a(\dot{\phi})$ has an extremum (minimum) at $\dot{\phi} =
\dot{\phi}_0$,  where  $a(\dot{\phi}_0)=a_0$, with
\begin{eqnarray}
\dot{\phi}_0 &=& \sqrt{\frac{A}{12|B|}}\,,\\
\nonumber\\
a_0 &=&
\left(\frac{12|B|}{A}\right)^{1/6}\left[\frac{3\Pi_{\phi}}{2A}\right]^{1/3}\,.
\end{eqnarray}

Also from Eq.~(\ref{Cons1a}), we obtain that in this region $a$
diverges when $\dot{\phi}$ approach to zero or to
$\sqrt{\frac{A}{4|B|}}$.

Therefore, we can note that a smaller scale factor than $a_0$ is out
of the range where the scale factor is defined for the physical
solutions.

\begin{figure}[h]
\begin{center}
\includegraphics[width=3.5in,angle=0,clip=true]{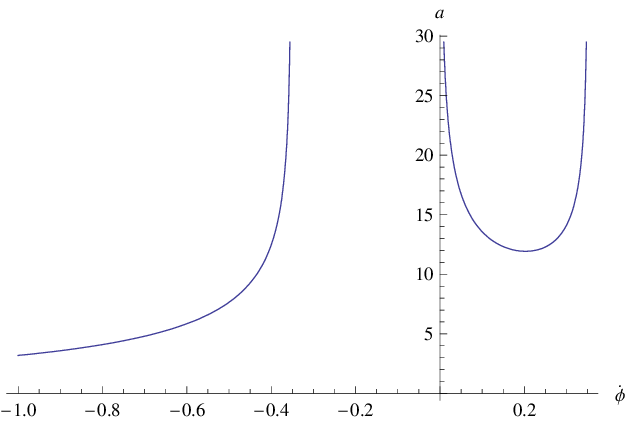}
\caption{From Eq.~(\ref{Cons1}), the scale factor $a$ as a function
of $\dot{\phi}$, when it is consider $\Pi_\phi =
113.41$.\label{Tres-Sol}}
\end{center}
\end{figure}

As an example, in Fig.~\ref{Tres-Sol} we have plotted
$a(\dot{\phi})$, where we have considered $B=-1$, $C=0.016$, $y
=0.505964$ and $\Pi_\phi = 113.41$.

From Eq.~(\ref{Cons1}) we can obtain $\dot{\phi}$ as a function of
$a$. We have three solutions:

\begin{eqnarray}
\dot{\phi}_1 &=& -\frac{a^3 A}{2\,3^{1/3} \left(9 a^6 B^2 \Pi_\phi
+\sqrt{3} \sqrt{a^{18} A^3 B^3+27 a^{12} B^4 \Pi_\phi
^2}\right)^{1/3}} \label{cu1}\\
\nonumber \\
&&+ \frac{\left(9 a^6 B^2 \Pi_\phi +\sqrt{3}\sqrt{a^{18} A^3 B^3+27
a^{12}B^4 \Pi_\phi^2}\right)^{1/3}}{2\,3^{2/3} a^3 B}\,,\nonumber \\
\nonumber \\
\dot{\phi}_2 &=& \frac{\left(1+i \sqrt{3}\right) a^3 A}{4 3^{1/3}
\left(9 a^6 B^2 \Pi_\phi +\sqrt{3} \sqrt{a^{18} A^3 B^3+27 a^{12}
B^4 \Pi_\phi^2}\right)^{1/3}} \label{cu2}\\
\nonumber \\
&&-\frac{\left(1-i \sqrt{3}\right) \left(9 a^6 B^2 \Pi_\phi
+\sqrt{3} \sqrt{a^{18} A^3 B^3+27 a^{12} B^4
\Pi_\phi^2}\right)^{1/3}}{4 3^{2/3}
a^3 B}\,,\nonumber \\ \nonumber\\
\dot{\phi}_3 &=& \frac{\left(1-i \sqrt{3}\right) a^3 A}{4 3^{1/3}
\left(9 a^6 B^2 \Pi_\phi +\sqrt{3} \sqrt{a^{18} A^3 B^3+27 a^{12}
B^4 \Pi_\phi^2}\right)^{1/3}}\label{cu3}\\
\nonumber \\
&&-\frac{\left(1+i \sqrt{3}\right) \left(9 a^6 B^2 \Pi_\phi
+\sqrt{3} \sqrt{a^{18} A^3 B^3+27 a^{12} B^4
\Pi_\phi^2}\right)^{1/3}}{4 3^{2/3} a^3 B}\,.\nonumber
\end{eqnarray}

\begin{figure}[h]
\begin{center}
\includegraphics[width=2.5in,angle=0,clip=true]{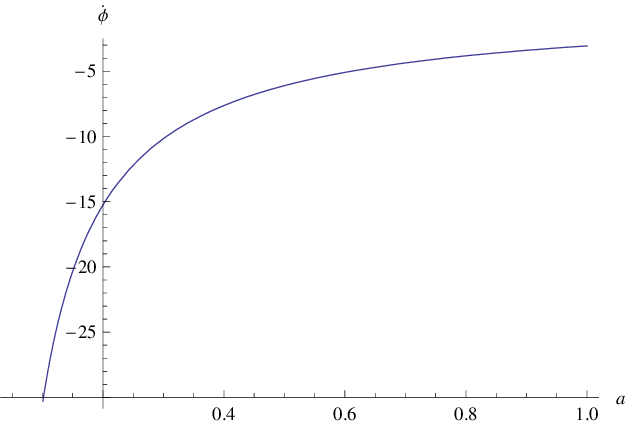}
\caption{From Eq.~(\ref{Cons1}), $\dot{\phi}_1$ as function of $a$
when it is consider $\Pi_\phi = 113.41$.\label{FSol1}}
\end{center}
\end{figure}

\begin{figure}[h]
\begin{center}
\includegraphics[width=2.5in,angle=0,clip=true]{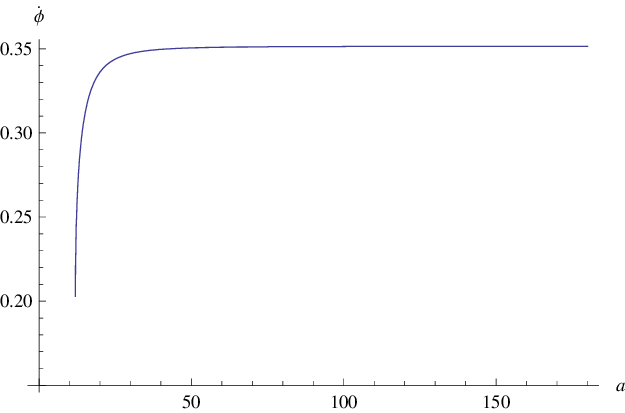}
\caption{From Eq.~(\ref{Cons1}), $\dot{\phi}_2$ as function of $a$
when it is consider $\Pi_\phi = 113.41$.\label{FSol2}}
\end{center}
\end{figure}

\begin{figure}[h]
\begin{center}
\includegraphics[width=2.5in,angle=0,clip=true]{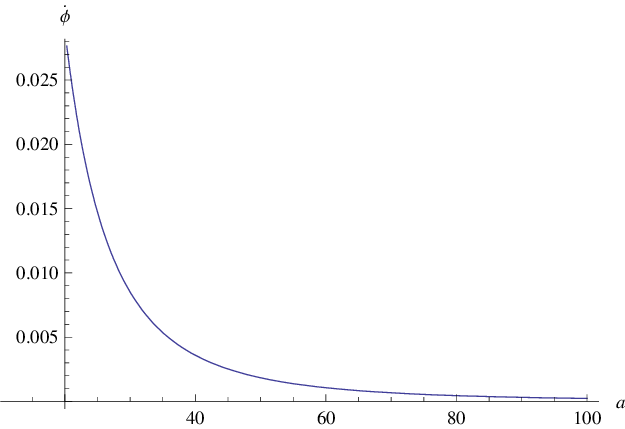}
\caption{From Eq.~(\ref{Cons1}), $\dot{\phi}_3$ as function of $a$
when it is consider $\Pi_\phi = 113.41$.\label{FSol3}}
\end{center}
\end{figure}

As an example we plot these solutions in Figs. (\ref{FSol1},
\ref{FSol2}, \ref{FSol3}), where we have used the values of
Ref.~\cite{yo} and $\Pi_\phi = 113.41$. We can note that the plots
are fully consistent with Fig.~\ref{Tres-Sol}.

The classical theory which describe this universe can be regarded as
a constrained dynamical system with a Hamiltonian

\begin{equation}
{\cal H} = -\frac{G}{3\pi\,a}\left(p^2_a + U(a)\right),
\end{equation}

where

\begin{equation}
p_a = - \frac{3\pi}{2G}\,a\,\dot{a},
\end{equation}

is the momentum conjugate to $a$, and $U(a)$ corresponds to the
effective potential given by

\begin{equation}\label{WD_Pot}
U(a) = \left(\frac{3\pi}{2G}\right) a^2\left(1 - \frac{8\pi G}{3}\,
a^2\, \rho(a)\right),
\end{equation}
where we have written $\rho$ as a function of $a$. It is possible to
do that by using the solutions Eqs.(\ref{cu1}, \ref{cu2}, \ref{cu3})
and Eq.~(\ref{eq.density}).

The Hamiltonian constraint is ${\cal H} = 0$, from where we obtain
the Friedmann equation

\begin{equation}\label{Friedmann}
\left(\frac{\dot{a}}{a}\right)^2 = \frac{8\pi G}{3}\,\rho(a) -
\frac{1}{a^2}\,.
\end{equation}

Also, we can obtain the equation for $\dot{H}$ given by

\begin{equation}\label{DH}
\dot{H} = - 4\pi\,G(\rho + p) + \frac{1}{a^2}\,.
\end{equation}
One of the characteristic of the EU scenario is the period of
superinflation after de static regimen and before inflation where
($\dot{H})>0$, see \cite{Labrana:2013oca}.
From Eq.~(\ref{DH}), we  note that in the relevant solutions of our
model we do not need to violate the null energy condition, $\rho + p
>0$, in order to have an EU scenario (with a superinflationary
phase) and avoid the initial singularity, since we are considering a
closed universe.
This is different from what happens in models as
Ref.~\cite{Rubakov:2014jja}, which by the way shows that some
violations of the energy conditions can be consistent.

In the context of quantum theory, the universe could be described by
a wave function $\psi(a)$, the conjugate momentum $p_a$ becomes the
differential operator $-id/da$ and the constraint is replaced by the
Wheeler-DeWitt (WDW)  equation \cite{Will}
\begin{eqnarray}
{\cal H} \psi(a) = 0, \\
\left( - \frac{d^2}{da^2} - \frac{\beta}{a}\,\frac{d}{da} + U(a)
\right)\psi(a) = 0\,,
\end{eqnarray}
where we have used the minisuperspace approximation, which is
appropriated for our model where the universe is homogeneous
isotropic and closed during the ES regimen and therefore has a
single degree of freedom, the scale factor \cite{Vilenkin:1987kf}.
The parameter $\beta$ represents the ambiguity in the ordering of
the non-commuting factors $a$ and $p_a$ in the Hamiltonian. The
value of $\beta$ does not affect the wave function in the
semiclassical regimen, and usually in the study of semi-classical
stability of EU it is chosen to be zero, see \cite{Mithani:2011en,
Mithani:2012ii, Mithani:2014jva, Mithani:2014toa}.

\begin{figure}[h]
\begin{center}
\includegraphics[width=3in,angle=0,clip=true]{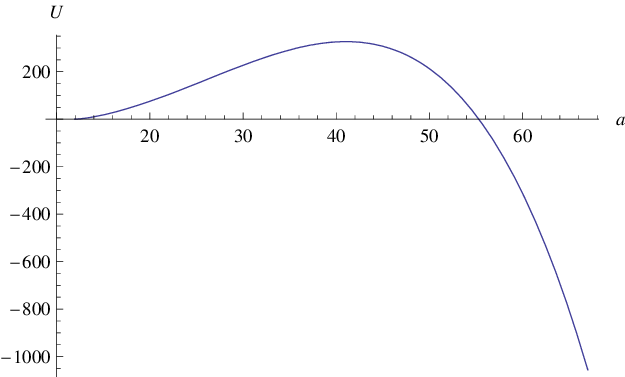}
\caption{Potential $U(a)$ for $a > a_0$. Here, we have used
Eq.~(\ref{cu2}).\label{FP1}}
\end{center}
\end{figure}

\begin{figure}[h]
\begin{center}
\includegraphics[width=3in,angle=0,clip=true]{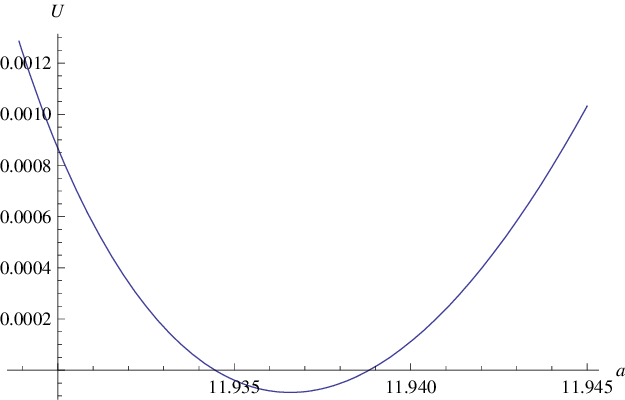}
\caption{Potential $U(a)$ for $a > a_0$ near the equilibrium point.
Here, we have used Eq.~(\ref{cu2}).\label{FP2}}
\end{center}
\end{figure}

\begin{figure}[h]
\begin{center}
\includegraphics[width=3in,angle=0,clip=true]{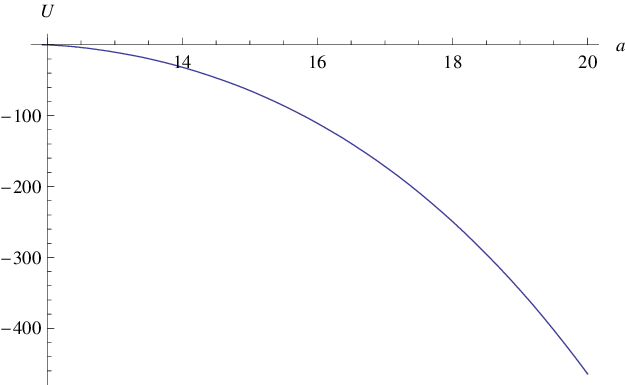}
\caption{Potential $U(a)$ for $a > a_0$. Here, we have used
Eq.~(\ref{cu3}).\label{FPVV}}
\end{center}
\end{figure}

In order to obtain the potential $U(a)$ for the case of the EU, we
have to select one of the solutions Eqs.(\ref{cu1}, \ref{cu2},
\ref{cu3}) which is related with the static and classically stable
solution. This classical solution was discussed in Ref.~\cite{yo}.
In this case this solution is Eq.~(\ref{cu2}). When we consider this
solution the potential $U(a)$ has a local minimum at $a=a^*$, where
$a^*$ was defined in Eq.~(\ref{consta}) and a local maximum at
$a=a'$, where
\begin{equation}\label{consta2}
a' = \sqrt{\left(\frac{3}{8\pi G}\right)\frac{12B}{A^2 + 24B\,C +
A\sqrt{A^2 + 12B\,C}}}\,\,.
\end{equation}
The nature of these two equilibrium points was discussed in
Ref.\cite{yo}, where it is shown that $a=a^*$ is an stable
equilibrium point and $a=a'$ is an unstable equilibrium point.
Then, the system is classically stable near the static solution, $a
\sim a^*$. There is a finite barrier which prevents the scale factor
to go from $a \simeq a^*$ to infinity and the potential is not well
defined for $a<a_0$ given the discussion above, this can be
interpreted as  a hard wall at $a=a_0$ for the potential $U(a)$.

As an example, in Fig.~\ref{FP1} it is plot the potential for the
values allowed for $a$, that is $a > a_0$. In Fig.~\ref{FP2} it is
plot the potential $U(a)$ near the static point (where also was
consider $a > a_0$).

Since a smaller scale factor than $a_0$ is out of the range where
the scale factor is defined for the physical solutions, we find that
the possible instability towards $a$ equal zero is not even a
logical possibility in this context. Nevertheless, we observe that
exist the possibility of tunneling through the finite barrier from
the static solution to an expanding universe, see Fig.~\ref{FP1}.
This is an interesting scenario to study in future works.

As an example, in Fig.~\ref{FPVV} it is show the potential for the
solution Eq.~(\ref{cu3}), we can note, as we expected, that in this
case there is not a equilibrium point as in Fig.~\ref{FP2}.

From the Friedmann equation, Eq.~(\ref{Friedmann}), and
Eq.~(\ref{Cons1}) we can note that solutions $\dot{\phi}_2$ and
$\dot{\phi}_3$ are not connected by the dynamics of the system.
Solution Eq.~(\ref{cu2}) satisfies $\dot{\phi}_2 >\dot{\phi}_0$ and
solution Eq.~(\ref{cu3}) satisfies $\dot{\phi}_3 <\dot{\phi}_0$, and
it is not possible to cross the line $\dot{\phi} = \dot{\phi}_0$. At
this respect, and by using Eq.(\ref{Cons1}), we can rewrite
Eq.~(\ref{Friedmann}) as the following equations for $\dot{\phi}$,

\begin{equation}
\ddot{\phi}^2 + V(\dot{\phi}) = 0\,,
\end{equation}
where

\begin{equation}\label{PotV}
V(\dot{\phi}) = \frac{(A\dot{\phi} + 4B\dot{\phi}^3)^2}{(A +
12B\dot{\phi}^2)^2}\,\left[ \frac{1}{\Pi_\phi^{2/3}}(A\dot{\phi} +
4B\dot{\phi}^3)^{2/3} -
\frac{\kappa}{3}\,\left(\frac{A}{2}\dot{\phi}^2 + 3B\dot{\phi}^4
+C\right) \right].
\end{equation}

Then, from Eq.~(\ref{PotV}) we note that the solutions
$\dot{\phi}_2$ and $\dot{\phi}_3$ are classically disconnected since
$V\rightarrow \infty$ at the value $\dot{\phi}= \dot{\phi}_0 =
\sqrt{\frac{A}{12|B|}}$. However, there is the possibility of
tunneling through this divergent barrier, see
\cite{Dittrich:1986ef}. In this case, the tunneling correspond to a
quantum tunneling from the static solution to an expanding universe
with initial values $a = a_0$.

\begin{figure}[h]
\begin{center}
\includegraphics[width=3in,angle=0,clip=true]{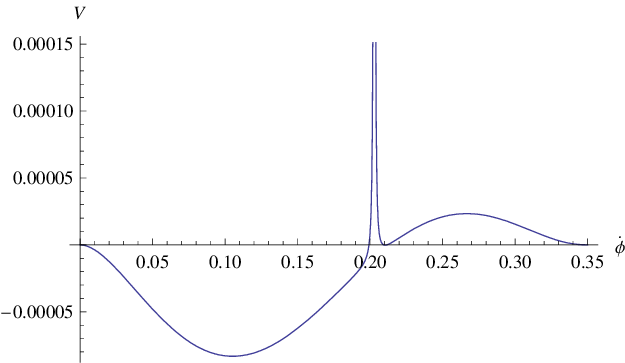}
\caption{Potential $V(\dot{\phi})$, for $\dot{\phi}
>0$.\label{FPP1}}
\end{center}
\end{figure}

As an example in Fig.~\ref{FPP1} it is shown the potential
$V(\dot{\phi})$, where we have used $\Pi_\phi = 113.41$ and the
values of Ref.~\cite{yo}. We can note that there is an infinite
barrier at $\dot{\phi} = \dot{\phi}_0 = 0.20$.

\begin{figure}[h]
\begin{center}
\includegraphics[width=3in,angle=0,clip=true]{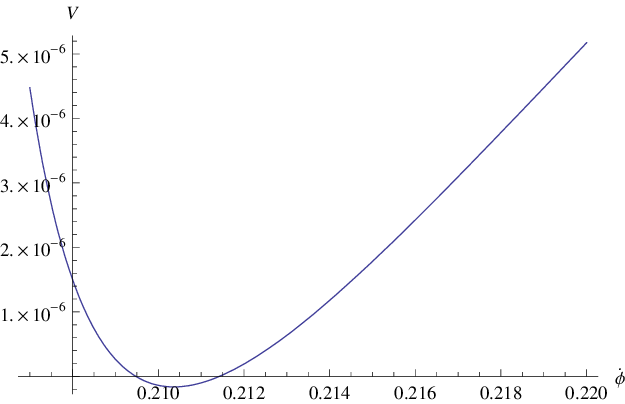}
\caption{Potential $V(\dot{\phi})$ near the equilibrium
point.\label{FPP2}}
\end{center}
\end{figure}

In Fig. \ref{FPP2} we show the dependence of the potential
$V(\dot{\phi})$ as a function of $\dot{\phi}$, near the equilibrium
point.

Therefore, we note that both tunnelings discussed above do not
correspond to a collapse to $a\rightarrow 0$, but a creation of an
expanding universe.

\section{Discussion and Conclusions}

It has been recently pointed out by Mithani-Vilenkin
\cite{Mithani:2014jva, Mithani:2011en, Mithani:2012ii,
Mithani:2014toa} that certain emergent universe scenarios which are
classically stable are nevertheless unstable semiclassically to
collapse. In this work, we shown that there is a class of emergent
universes derived from scale invariant two measures theories with
spontaneous symmetry breaking of the scale invariance, which can
have both classical stability and do not suffer the instability
pointed out by Mithani-Vilenkin towards collapse.
This stability is due to the presence of a symmetry in the "emergent
phase", which together with the non linearities of the theory, does
not allow the FLRW scale factor to be smaller that a certain minimum
$a_0$ in a certain protected region.

Since a smaller  scale factor than $a_0$ is out of the range where
it is defined for the physical solutions $\dot{\phi}_2$  and
$\dot{\phi}_3$ where $\rho$ is positive, we have found that the
possible instability towards a scale factor equal zero is not even a
logical possibility in this context. Therefore our model is free of
the instability towards collapse described in
Refs.~\cite{Mithani:2014jva, Mithani:2011en, Mithani:2012ii,
Mithani:2014toa}. The conserved quantity $\Pi_\phi \neq 0$ provides
in this case with a protection towards collapse to $a$ equal zero.

It is interesting to observe that exist the possibility of tunneling
through the finite barrier of the potentials $U(a)$ and
$V(\dot{\phi})$ from the static solution to an expanding universe,
but also there is the possibility of tunneling through the divergent
barrier of potentials $V(\dot{\phi})$, see Eq.~(\ref{PotV}). In this
case, the tunneling correspond to a quantum tunneling from the
static solution to an expanding universe with initial values $a =
a_0$.
We have noted that both tunnelings process, do not correspond to a
collapse to $a\rightarrow 0$, instead they correspond to a creation
of an expanding universe.
This is an interesting scenario to study in future works and
correspond to an alternative scheme for an emergent universe
scenario, similar to the one studied in Refs.~\cite{Labrana:2011np}.

In particular in this work we studied the model Ref.~\cite{yo}, but
the results obtained in this work can also be applied to  models
studied in Refs.~\cite{prev}-\cite{Guendelman:2014bva}, which
present similar symmetries as the model in Ref.~\cite{yo}.

We should mention that, we have considered a closed universe, where
the contribution of the curvature term is relevant before the
inflationary period. Nevertheless, they are the possibility of
contrast the EU with observation by studying the superinflationary
period of these models. As it was reported in
Ref.~\cite{Labrana:2013oca}, during the superinflationary period,
the EU scenario produces a suppression of the CMB anisotropies at
large scale which could be responsible for the observed lack of
power at large angular scales of the CMB. We hope to be able to
analyze this suppression and also submitting our model to further
test such as CMB temperature anisotropies and density perturbations.
This will be the subject of a future work.

\section{Acknowledgements}
The authors dedicate this article to the memory of Professor Sergio
del Campo (R.I.P.).

We thank Professor Alexander Vilenkin for suggesting and encouraging
us to study the problem of the quantum stability of the emergent
universe scenario and for multiple discussions on this subject.
This work was supported by Comisi\'on Nacional de Ciencias y
Tecnolog\'{\i}a through FONDECYT  Grants 1110230 (SdC), 1130628
(RH). Also it was supported by Pontificia Universidad Cat\'olica de
Valpara\'{\i}so  through grants 123.787-2007 (SdC) and 123724 (RH).
One of us (E.I.G) would like to thank the astrophysics and cosmology
group at the Pontificia Universidad Cat\'olica de Valpara\'{\i}so
and the Frankfurt Institute of Advanced Studies of Frankfurt
University for hospitality. P. L. is supported by Direcci\'on de
Investigaci\'on de la Universidad del B\'{\i}o-B\'{\i}o through
Grants N$^0$ 166907 2/R, and GI 150407/VC.


\section*{Appendix A: Review of the TMT theories}\label{app1}


The TMT is a generally coordinate invariant theory, where  the
action has to be of the form\cite{GK3}-\cite{GK8}
\begin{equation}
    S = \int L_{1}\Phi d^{4}x +\int L_{2}\sqrt{-g}d^{4}x,
\label{S}
\end{equation}
including two Lagrangians $ L_{1}$ and $L_{2}$ and two measures of
integration $\sqrt{-g}$ and $\Phi$. One is the usual measure of
integration $\sqrt{-g}$ in the 4-dimensional space-time manifold
equipped with the metric
 $g_{\mu\nu}$. The other  is the new measure of integration $\Phi$ in the same
4-dimensional space-time manifold. The measure  $\Phi$ being  a
scalar density and a total derivative (see Ref.\cite{Mstring}) may
be defined by means of  four scalar fields $\varphi_{a}$
($a=1,2,3,4$),
\begin{equation}
\Phi
=\varepsilon^{\mu\nu\alpha\beta}\varepsilon_{abcd}\partial_{\mu}\varphi_{a}
\partial_{\nu}\varphi_{b}\partial_{\alpha}\varphi_{c}
\partial_{\beta}\varphi_{d}.
\label{Phi}
\end{equation}

It is assumed that the Lagrangian densities $ L_{1}$ and $L_{2}$ are
functions of all matter fields, the dilaton field, the metric, the
connection
 but not of the
"measure fields" ($\varphi_{a}$ ). In such a case, i.e. when the
measure fields  enter in the theory only via the measure $\Phi$,
  the action (\ref{S}) possesses
an infinite dimensional symmetry. In the case given by
Eq.(\ref{Phi}) these symmetry transformations have the form
$\varphi_{a}\rightarrow\varphi_{a}+f_{a}(L_{1})$, where
$f_{a}(L_{1})$ are arbitrary functions of  $L_{1}$ (see details in
Ref.\cite{GK3}).

In this theory, we assume that all fields, including also the
metric, connection and the {\it measure fields} are independent
dynamical variables. All the relations among  them are results of
the equations of motion. In particular, the independence of the
metric and the connection means that we proceed in the first order
formalism and the relation between connection and metric is not
necessarily according to Riemannian geometry.

Varying the measure fields $\varphi_{a}$, we get
$B^{\mu}_{a}\partial_{\mu}L_{1}=0 $ where
$B^{\mu}_{a}=\varepsilon^{\mu\nu\alpha\beta}\varepsilon_{abcd}
\partial_{\nu}\varphi_{b}\partial_{\alpha}\varphi_{c}
\partial_{\beta}\varphi_{d}.\label{varphiB}$
Since $Det (B^{\mu}_{a}) = \frac{4^{-4}}{4!}\Phi^{3}$ it follows
that if $\Phi\neq 0$,
\begin{equation}
 L_{1}=sM^{4} =const,
\label{varphi}
\end{equation}
where $s=\pm 1$ and $M$ is a constant of integration with the
dimension of mass.

We proceed now  to discuss the question of scale invariance in the
context of TMT. A dilaton field $\phi$ allows to realize a
spontaneously broken global scale invariance\cite{G1}. We postulate
that the theory is invariant under the global scale transformations:
\begin{equation}
    g_{\mu\nu}\rightarrow e^{\theta }g_{\mu\nu}, \quad
\Gamma^{\mu}_{\alpha\beta}\rightarrow \Gamma^{\mu}_{\alpha\beta},
\quad \varphi_{a}\rightarrow \lambda_{ab}\varphi_{b}\quad
\text{where} \quad \det(\lambda_{ab})=e^{2\theta}, \quad
\phi\rightarrow \phi-\frac{M_{p}}{\alpha}\theta . \label{st}
\end{equation}

We choose an action which, except for the modification of the
general structure caused by the basic assumptions of TMT, {\it does
not contain any exotic terms and  fields} as like as in the
conventional formulation of the minimally coupled scalar-gravity
system. Keeping the general structure (\ref{S}), it is convenient to
represent the underlying action of our model in the following form
\cite{GKKess}:

\begin{eqnarray}
S &=&\int d^{4}x e^{\alpha\phi /M_{p}}
\Big[-\frac{1}{2\,\kappa}R(\Gamma ,g)(\Phi +b_{g}\sqrt{-g})+(\Phi
+b_{\phi}\sqrt{-g})\frac{1}{2}g^{\mu\nu}\phi_{,\mu}\phi_{,\nu}
\\
&-& e^{\alpha\phi /M_{p}}\left(\Phi V_{1}
+\sqrt{-g}V_{2}\right)\Big].\nonumber
 \label{totaction}
\end{eqnarray}
We use $\kappa =8\pi/M_p^2 $ where $M_p$ is the four-dimensional
Planck mass. In the equations of motion following  from this action,
we change the metric to the new one
\begin{equation}
\tilde{g}_{\mu\nu}=e^{\alpha\phi/M_{p}}(\zeta +b_{g})g_{\mu\nu},
\label{ct}
\end{equation}
where $\zeta \equiv\frac{\Phi}{\sqrt{-g}} \label{zeta}$. The
conformal metric $\tilde{g}_{\mu\nu}$  represents  the "Einstein
frame", since the connection  becomes Riemannian. Notice that
$\tilde{g}_{\mu\nu}$ is invariant under the scale transformations
(\ref{st}). After the change of variables  to the Einstein frame the
gravitational equations take the standard GR form
\begin{equation}
G_{\mu\nu}(\tilde{g}_{\alpha\beta})=\kappa\,T_{\mu\nu}^{eff},
 \label{gef}
\end{equation}
where  $G_{\mu\nu}(\tilde{g}_{\alpha\beta})$ is the Einstein tensor.
The energy-momentum tensor, $T_{\mu\nu}^{eff}$, becomes
\begin{eqnarray}
T_{\mu\nu}^{eff}&=&\frac{\zeta +b_{\phi}}{\zeta +b_{g}}
\left(\phi_{,\mu}\phi_{,\nu}-\frac{1}{2}
\tilde{g}_{\mu\nu}\tilde{g}^{\alpha\beta}\phi_{,\alpha}\phi_{,\beta}\right)
-\tilde{g}_{\mu\nu}\frac{b_{g}-b_{\phi}}{2(\zeta +b_{g})}
\tilde{g}^{\alpha\beta}\phi_{,\alpha}\phi_{,\beta}
+\tilde{g}_{\mu\nu}V_{eff}(\phi;\zeta,M),
 \label{Tmn}
\end{eqnarray}
where the function $V_{eff}(\phi;\zeta,M)$ is defined as following:
\begin{equation}
V_{eff}(\phi;\zeta ,M)=
\frac{b_{g}\left[sM^{4}e^{-2\alpha\phi/M_{p}}+V_{1}\right]
-V_{2}}{(\zeta +b_{g})^{2}}. \label{Veff1}
\end{equation}

The scalar field $\zeta$  is determined by the consistency of
(\ref{gef}) with (\ref{varphi}), which lead to the constraint
\begin{eqnarray}
&&(b_{g}-\zeta)\left[sM^{4}e^{-2\alpha\phi/M_{p}}+
V_{1}\right]-2V_{2}-\delta\cdot b_{g}(\zeta +b_{g})Z
=0,\label{constraint2}
\end{eqnarray}
where
$Z\equiv\frac{1}{2}\tilde{g}^{\alpha\beta}\phi_{,\alpha}\phi_{,\beta}$
and $\delta =\frac{b_{g}-b_{\phi}}{b_{g}}$.

The effective energy-momentum tensor (\ref{Tmn}) can be represented
in a form of that of  a perfect fluid $T_{\mu\nu}^{eff}=(\rho
+p)u_{\mu}u_{\nu}-p\tilde{g}_{\mu\nu}$, where
$u_{\mu}=\frac{\phi_{,\mu}}{(2Z)^{1/2}}$ with the following energy
and pressure densities resulting from Eqs.(\ref{Tmn}) and
(\ref{Veff1}) after inserting the solution $\zeta =\zeta(\phi,Z;M)$
of Eq.(\ref{constraint2})
\begin{equation}
\rho(\phi,Z;M) =Z+ \frac{(sM^{4}e^{-2\alpha\phi/M_{p}}+V_{1})^{2}-
2\delta b_{g}(sM^{4}e^{-2\alpha\phi/M_{p}}+V_{1})Z -3\delta^{2}
b_{g}^{2}Z^2}{4[b_{g}(sM^{4}e^{-2\alpha\phi/M_{p}}+V_{1})-V_{2}]},
\label{rho1}
\end{equation}
and
\begin{equation}
p(\phi,Z;M) =Z- \frac{\left(sM^{4}e^{-2\alpha\phi/M_{p}}+V_{1}+
\delta b_{g}Z\right)^2}
{4[b_{g}(sM^{4}e^{-2\alpha\phi/M_{p}}+V_{1})-V_{2}]}. \label{p1}
\end{equation}
Notice that if $s$ and $V_{1}$ have different signs one obtains
without fine tuning a vacuum state with zero energy density.
In this work we will consider a scenario where the scalar field is
moving in the extreme left region $\phi \rightarrow - \infty$, then,
in this case $\alpha <0$. The constants of this model are subject to
the observational constrains and stability conditions
Eqs.~(\ref{ES1}-\ref{ES3}) studied in Ref.~\cite{yo}.


\section*{Appendix B: Case $\Pi_\phi < 0$}\label{app2}

We consider the case $\Pi_\phi < 0$.
From conservation equation (\ref{Cons1}), we can write $a$ as a
function of $\dot{\phi}$
\begin{equation}\label{ACons1a}
a(\dot{\phi}) =
\left(\frac{\Pi_\phi}{A\dot{\phi}+4\,B\dot{\phi}^3}\right)^{1/3} .
\end{equation}

We can note that in this case $-\sqrt{\frac{A}{4|B|}}< \dot{\phi}
<0$ or $\sqrt{\frac{A}{4|B|}}< \dot{\phi} < \infty$ in order to
satisfied $\Pi_\phi < 0$.
When $\dot{\phi}$ is in the second region $a(\dot{\phi})$ is a
function which approach to zero when $\dot{\phi} \rightarrow \infty$
and diverges when $\dot{\phi} \rightarrow \sqrt{\frac{A}{4|B|}}$.
But in this region $\rho$ becomes negative see
Eq.~(\ref{eq.density}), then we are not interested in this case.

On the other hand, when $\dot{\phi}$ is in the first region,
$a(\dot{\phi})$ has a minimum at $\dot{\phi} = \dot{\phi}_0$, where
$a(\dot{\phi}_0)=a_0$, with
\begin{eqnarray}
\dot{\phi}_0 &=& -\sqrt{\frac{A}{12|B|}}\,,\\
\nonumber\\
a_0 &=&
\left(\frac{12|B|}{A}\right)^{1/6}\left[\frac{3|\Pi_{\phi}|}{2A}\right]^{1/3}\,.
\end{eqnarray}

Also from Eq.~(\ref{ACons1a}), we obtain that in this region $a$
diverges when $\dot{\phi}$ approach to zero or to
$-\sqrt{\frac{A}{4|B|}}$.

Therefore, we can note that a smaller scale factor than $a_0$ is out
of the range where the scale factor is defined for the physical
solutions.

\begin{figure}[h]
\begin{center}
\includegraphics[width=3.5in,angle=0,clip=true]{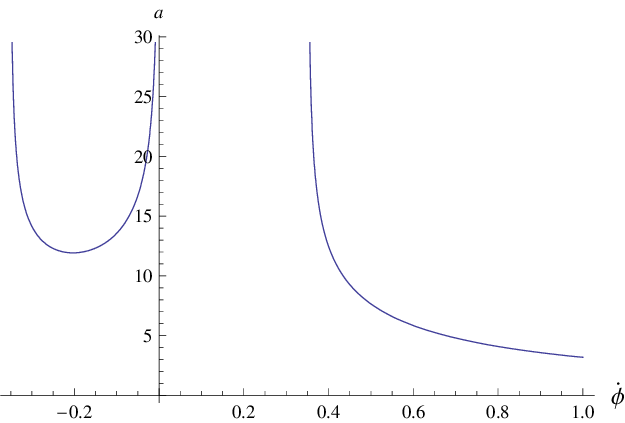}
\caption{From Eq.~(\ref{Cons1}), the scale factor $a$ as a function
of $\dot{\phi}$, when it is consider $\Pi_\phi =
-113.41$.\label{A-Tres-Sol}}
\end{center}
\end{figure}

As an example, in Fig.~\ref{A-Tres-Sol} we have plotted
$a(\dot{\phi})$, where we have considered $B=-1$, $C=0.016$, $y
=0.505964$ and $\Pi_\phi = -113.41$.

\end{document}